\documentclass[conference]{IEEEtran}
\IEEEoverridecommandlockouts
\usepackage{cite}
\usepackage{balance}
\usepackage{amsmath,amssymb,amsfonts}
\usepackage{graphicx}
\usepackage{textcomp}
\usepackage{xcolor}
\def\BibTeX{{\rm B\kern-.05em{\sc i\kern-.025em b}\kern-.08em
    T\kern-.1667em\lower.7ex\hbox{E}\kern-.125emX}}

\makeatletter
\renewcommand{\fnum@figure}{Fig. \thefigure}
\makeatother

\usepackage[left=1.608cm,right=1.608cm,top=1.8cm]{geometry}
\usepackage{tikz}
\usetikzlibrary{shapes,arrows,positioning,decorations.pathreplacing,calc,fit}
\usepackage{algorithm}
\usepackage{algpseudocode}
\usepackage{colortbl}    
\usepackage{xcolor}
\usepackage{tikz}
\usepackage{diagbox}     
\usepackage{array}       
\usepackage{changepage} 
\usepackage{siunitx}
\usepackage{booktabs}

\usepackage{pgfplots}
\usepackage{subcaption}
\pgfplotsset{compat=1.18}
\usepackage{pgfplotstable}
\usetikzlibrary{positioning,matrix, arrows.meta, fit,  shapes.arrows}
\usepackage{multirow}
\usepackage{pgfplots, pgfplotstable}
\usepgfplotslibrary{statistics}
\usetikzlibrary{positioning, matrix, arrows.meta, shapes.arrows}

\usepackage[acronym]{glossaries}
\newacronym{OFDM}{OFDM}{orthogonal frequency division multiplexing}
\newacronym{CP}{CP}{cyclic prefix}
\newacronym{JCAS}{JCAS}{joint communication and  radar sensing}
\newacronym{QAM}{QAM}{quadrature amplitude modulation}
\newacronym{MUSIC}{MUSIC}{multiple signal classification}
\newacronym{ESPRIT}{ESPRIT}{estimation of signal parameters via rotational invariance technique}
\newacronym{CS}{CS}{compressive sensing}
\newacronym{FPS}{FPS}{Fourier projection-slice}
\newacronym{LMMSE}{LMMSE}{linear minimum mean square error}
\newacronym{LS}{LS}{least square }
\newacronym{MMSE}{MMSE}{minimum mean square error}
\newacronym{CE}{CE}{channel estimation}
\newacronym{RMSE}{RMSE}{root mean square error}
\newacronym{TDD}{TDD}{Time Division Duplexing}
\newacronym{FDD}{FDD}{Frequency Division Duplexing}
\newacronym{UL}{UL}{Uplink}
\newacronym{DL}{DL}{deep learning}
\newacronym{BS}{BS}{base station}
\newacronym{MRC}{MRC}{Maximum Ratio Combining}
\newacronym{CSI}{CSI}{channel state information}
\newacronym{ACI}{ACI}{adjacent channel interference}
\newacronym[longplural={User Equipments}]{UE}{UE}{User Equipment}
\newacronym{3GPP}{3GPP}{3rd generation partnership project }
\newacronym{LTE}{LTE}{long term evolution}
\newacronym{GT}{GT}{ground-truth}
\newacronym{SFT}{SFT}{sparse Fourier transform}
\newacronym{ED}{ED}{Energy Detection}
\newacronym{ASK}{ASK}{Amplitude-shift keying}
\newacronym{PSK}{PSK}{Phase-shift keying}
\newacronym{ZCP}{ZCP}{Zadoff-Chu precoding}
\newacronym{ISM}{ISM}{industrial scientific and medical}
\newacronym{SNR}{SNR}{signal-to-noise ratio}
\newacronym{IDFT}{IDFT}{inverse discrete Fourier transform }
\newacronym{DFT}{DFT}{discrete Fourier transform }
\newacronym{IFFT}{IFFT}{inverse fast Fourier transform }
\newacronym{FFT}{FFT}{fast Fourier transform }
\newacronym{ISI}{ISI}{inter symbol interference}
\newacronym{ICI}{ICI}{inter carrier interference}
\newacronym{MIMO}{MIMO}{multiple input multiple output}
\newacronym{PAPR}{PAPR}{peak-to-average power ratio}
\newacronym{GLRT}{GLRT}{Generalized Likelihood Ratio Test}
\newacronym{i.i.d}{i.i.d}{independent and identically distributed}
\newacronym{PSNR}{PSNR}{peak signal to noise ratio}
\newacronym{AWGN}{AWGN}{additive white Gaussian noise}
\newacronym{ZCT}{ZCT}{Zaddof-Chu transform}
\newacronym{HPA}{HPA}{high power amplifier}
\newacronym{RFI}{RFI}{Radio Frequency Interferences}
\newacronym{2D}{2D}{two-dimensional}
\newacronym{1D}{1D}{one-dimensional}
\newacronym{MSE}{MSE}{mean square error}
\newacronym{BER}{BER}{bit error rate}
\newacronym{DCP}{DCP}{disciplined convex programming}

\newacronym{ISAC}{ISAC}{integrated sensing and communication}
\newacronym{ULA}{ULA}{uniform linear array}
\newacronym{HPSD}{HPSD}{Hermittian positive semidefinite}
\newacronym{SINR}{SINR}{signal to interference plus noise ratio}
\newacronym{SCA}{SCA}{successive convex approximation}
\newacronym{SDP}{SDP}{semidefinite programming}
\newacronym{CDF}{CDF}{cumulative distribution function}
\newacronym{NOMA}{NOMA}{non-orthogonal multiple access}
\newacronym{AoA}{AoA}{angle-of-arrival}
\newacronym{IoT}{IoT}{internet of things}
\newacronym{NLoS}{NLoS}{non-line-of-sight}
\newacronym{LoS}{LoS}{line-of-sight}
\newacronym{CFO}{CFO}{carrier frequency offset}
\newacronym{CPO}{CPO}{carrier phase offset}
\newacronym{SFO}{SFO}{sampling frequency offset}
\newacronym{STO}{STO}{sampling time offset}
\newacronym{ToF}{ToF}{time-of-flight}
\newacronym{SVM}{SVM}{support vector machine}
\newacronym{LR}{LR}{logistic regression}
\newacronym{RF}{RF}{random forest}
\newacronym{RSSI}{RSSI}{received signal strength indicator}
\newacronym{RSS}{RSS}{received signal strength}
\newacronym{MISO}{MISO}{multiple input single output}
\newacronym{SIMO}{SIMO}{single input multiple output}
\newacronym{DAC}{DAC}{digital-to-analog converter}
\newacronym{ADC}{ADC}{analog-to-digital converter}
\newacronym{CIR}{CIR}{channel impulse response}
\newacronym{PCA}{PCA}{principal component analysis}
\newacronym{PLA}{PLA}{physical layer authentication}
\newacronym{TDoA}{TDoA}{time difference of arrival}
\newacronym{ToA}{ToA}{time of arrival}
\newacronym{IRS}{IRS}{intelligent reflecting surface}
\newacronym{GMM}{GMM}{Gaussian mixture model}
\newacronym{A-GMM}{A-GMM}{Adaptive Gaussian mixture model}
\newacronym{LOF}{LOF}{local outlier factor}
\newacronym{k-NN}{k-NN}{k-nearest neighbors}
\newacronym{IQ}{IQ}{in-phase and quadrature}
\newacronym{FAR}{FAR}{false alarm rate}
\newacronym{DR}{DR}{detection rate}
\newacronym{MAE}{MAE}{mean absolute error}  

\begin{document}
  
\title{Physical Layer Anomaly Detection via Adaptive Gaussian Mixture Model
}



\author{
\IEEEauthorblockN{
    Mamady  Delamou\IEEEauthorrefmark{1}, 
    Luan  Chen \IEEEauthorrefmark{1},
    Arsenia  Chorti \IEEEauthorrefmark{1}\IEEEauthorrefmark{2}\\
}
    \IEEEauthorblockA{\IEEEauthorrefmark{1}ETIS UMR 8051, CY Cergy Paris Université, ENSEA, CNRS, Cergy, France,\\ 
    }
    \IEEEauthorblockA{\IEEEauthorrefmark{2} Barkhausen Institut, Dresden, Germany,\\ 
} 

{\{mamady.delamou, luan.chen\}@ensea.fr}}

\maketitle
\begin{abstract}
Physical-layer spoofing attacks pose a critical threat to wireless localization systems by manipulating channel state information (CSI). Detecting such attacks becomes particularly challenging when the attacker and the legitimate node are spatially proximate, as their CSI characteristics become increasingly similar. To address this challenge in indoor scenarios, we propose an adaptive Gaussian mixture model (GMM)-based detection framework that learns legitimate CSI patterns and detects spatial inconsistency through statistical anomaly detection. Using a massive multiple-input multiple-output (MIMO) experimental dataset with 64 uniform linear array (ULA) antenna and 5 mm spatial resolution, we evaluated our detector against CSI spoofing attacks with spatial displacements ranging from 50 mm to 450 mm. The adaptive GMM achieves 95-98\% detection rate for displacements above 150 mm while maintaining 1\% false alarm rate. We benchmark our approach against conventional GMM, local outlier factor (LOF), one-class support vector machine (SVM) and k-nearest neighbors (k-NN) classifiers, demonstrating overall superior detection performance of our proposed classifier.
\end{abstract}


\section{Introduction}\label{sec:introduction}

Wireless localization systems serve as fundamental infrastructure for numerous critical applications including indoor navigation, asset tracking, autonomous vehicles and emergency services \cite{10986849, Zafari2017ASO}.
The proliferation of internet of things and the emergence of ultra-reliable low-latency communication requirements in 5G and beyond have further amplified the importance of accurate and secure localization \cite{10912068}.
These systems typically rely on physical layer metrics such as  time of arrival, time difference of arrival, \gls{AoA}, \gls{ToF}, and \gls{RSS} to estimate user positions with different degrees of accuracy. 
However, the fundamental reliance on the characteristics of electromagnetic signal propagation makes these systems inherently vulnerable to physical layer attacks, where adversaries can exploit the wireless channel to compromise positioning integrity \cite{10669078, 9828797, pham2025leveraging}. Such vulnerabilities pose significant security risks in applications ranging from access control systems to vehicular communication networks \cite{10829892, 11102160}.

Ensuring the security of wireless localization systems requires robust mechanisms to distinguish legitimate transmission from adversarial signal. The challenge lies in the fact that both legitimate and corrupted signals are embedded in noisy physical layer observations, making discrimination difficult \cite{8304641}. Physical layer measurements are inherently subject to various sources of distortion, including multipath propagation \cite{9338751}, hardware imperfections, and environmental variations, which can mask or obscure attack signatures. Furthermore, the multivariate nature of localization measurements, where multiple metrics such as \gls{AoA}, \gls{ToF} and \gls{RSS} are jointly observed, introduces complex dependencies that must be properly modeled for effective attack detection \cite{8911771}.

Machine learning-based methods for anomaly detection offer a principled approach to this problem by learning the characteristics of the signal directly from the measurement data \cite{9439459, 8911771}. By modeling the joint distribution of multiple metrics, these methods can capture complex patterns. On this basis, we propose an adaptive \gls{GMM}-based detection framework that dynamically adjusts detection thresholds based on local measurement statistics, enabling robust attack detection. To validate the effectiveness of our approach, we benchmark it against three established anomaly detection algorithms such as \gls{LOF}, one-class \gls{SVM}, and \gls{k-NN}, demonstrating its superior performance.

Thoroughly, our framework leverages the joint distribution of \gls{AoA} and \gls{RSS} measurements to capture the spatial constraints of legitimate user motion patterns. The effectiveness of this approach relies on the fundamental observation that legitimate localization measurements, despite being subject to estimation noise, exhibit structured patterns in the joint feature space. Physical constraints on user mobility create natural correlations between consecutive measurements that attackers cannot easily replicate without knowing the true trajectory of the legitimate user \cite{10669078}. When an adversary spoofs \gls{CSI}, these attacks necessarily violate the learned patterns, creating detectable anomalies in the feature space. By modeling these patterns probabilistically with adaptive thresholding, we enable robust detection without requiring prior knowledge of specific attack strategies \cite{9828797}.



The main contributions of this work are summarized as follows:

\begin{enumerate}

\item We calibrate the \gls{CSI} matrices to mitigate hardware imperfections and frequency-dependent errors in the measurements, ensuring high-quality input features for the detection framework.

\item We propose an adaptive \gls{GMM}-based detection framework that jointly analyzes \gls{AoA} and \gls{RSS} measurements. The method formulates location spoofing attack detection as an anomaly detection task by exploiting violations of spatial consistency patterns inherent in legitimate user motion.

\item We compare our proposed adaptive \gls{GMM} against \gls{LOF}, one-class \gls{SVM} and \gls{k-NN}, demonstrating that it achieves 95-98\% detection rate for spatial displacements ranging from 150 mm.
\end{enumerate}

The remainder of the paper is organized as follows: In Section \ref{sec:system_model}, we introduce the system model and attack scenarios. Subsequently, in Section \ref{sec:gmm_detection}, we detail our proposed an adaptive \gls{GMM} framework. Section \ref{sec:experiment_results}, presents the experimental evaluation and comparative analysis, and finally, in Section \ref{sec:conclusion}, we conclude and highlight future directions.

\section{System Model And Problem Statement}\label{sec:system_model}

\begin{figure}[htbp]
  \centering
  \includegraphics[width=\linewidth]{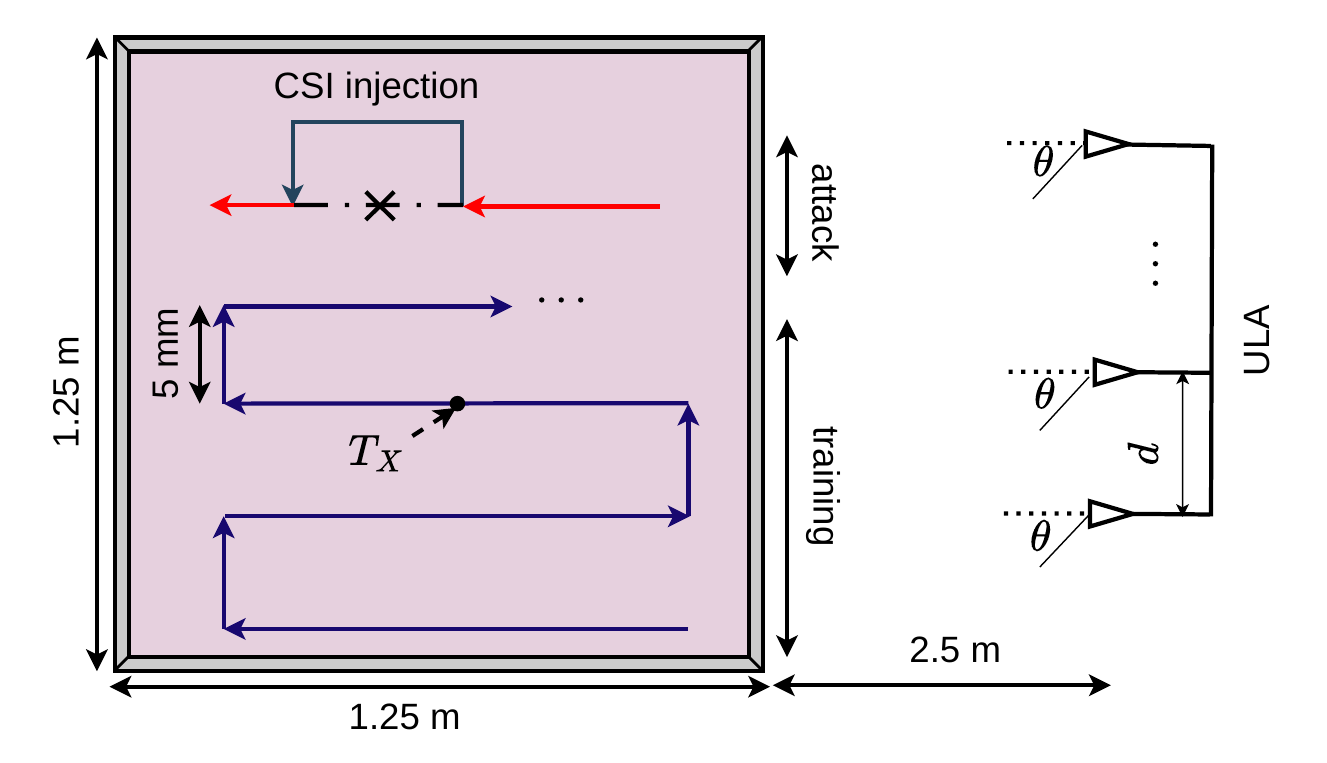}
  \caption{Massive MIMO measurement campaign of CSI collection using a \gls{ULA}. The system consists of learning or training step (blue trajectories) and testing or attack step (red trajectory).}
  \label{fig:myfigure}
\end{figure}
\setlength{\textfloatsep}{0pt}
As shown in Fig.\ref{fig:myfigure}, we consider a massive MIMO localization system operating in an ultra-dense indoor \gls{LoS} environment \cite{9535488}. The base station is equipped with a \gls{ULA} with inter-antenna spacing $d = 0.07$ m at the operating carrier frequency $f_\text{c} = 2.61$ GHz. The system operates with a bandwidth $B = 20$ MHz divided into $N_\text{sc} = 100$ \gls{OFDM} subcarriers, yielding a subcarrier spacing of $\Delta f = 200$ kHz. User equipment consists of single-antenna devices transmitting at power $P_\text{tx} = 15$ dBm. The wavelength at the carrier frequency is $\lambda = c/f_\text{c} \approx 0.115$ m, where $c$ denotes the speed of light.

The wireless channel between the user equipment ($T_\text{X}$) and the massive MIMO base station ($R_\text{X}$) at time $t$ is represented by the channel matrix $\mathbf{H}(t) \in \mathbb{C}^{N_\text{a} \times N_\text{sc}}$, where $N_\text{a}$ denotes the number of antennas and $N_\text{sc}$ represents the number of subcarriers, with $ 0 \leq a \leq N_\text{a}-1$ and $0 \leq s \leq N_\text{sc}-1$. Under \gls{LoS} propagation conditions, the channel matrix can be decomposed as:
$\mathbf{H}(t) = \alpha(t) \mathbf{a}(\theta) \otimes \mathbf{f}(\tau) + \mathbf{N}(t)$, where $\alpha(t)$ represents the complex channel gain that includes path loss and small-scale fading, $\mathbf{a}(\theta) \in \mathbb{C}^{N_\text{a}}$ denotes the spatial steering vector determined by \gls{AoA} $\theta$, $\mathbf{f}(\tau) \in \mathbb{C}^{N_\text{sc}}$ represents the frequency steering vector determined by propagation delay $\tau$, $\mathbf{N}(t)$ captures additive white Gaussian noise and $\otimes$ is the Kronecker product.

The spatial steering vector for \gls{ULA} with antenna spacing $d$ is expressed as follows:
\begin{equation}
\mathbf{a}(\theta) = \left[1, e^{-j\frac{2\pi d}{\lambda}\sin\theta}, \ldots, e^{-j\frac{2\pi d(N_\text{a}-1)}{\lambda}\sin\theta}\right]^T.
\end{equation}

The frequency steering vector exploits the frequency selectivity of the propagation delay across OFDM subcarriers, which can be expressed as follows:
\begin{equation}
\mathbf{f}(\tau) = \left[1, e^{-j2\pi\Delta f\tau}, \ldots, e^{-j2\pi(N_\text{sc}-1)\Delta f\tau}\right]^T.
\end{equation}



We consider an attacker who can inject pre-recorded \gls{CSI} measurements from a different spatial location into the localization system. This adversary pre-collects \gls{CSI} measurements from different locations and transmits them to spoof their position. Formally, let $\mathbf{H}_{\text{legit}}(t)$ denote the legitimate channel matrix corresponding to the user's true trajectory and let $\mathbf{H}_{\text{attack}}(t)$ denote the spoofed channel matrix from a different trajectory. During the attack $t \in [t_{\text{start}}, t_{\text{end}}]$, the system observes spoofed channels instead of the legitimate ones.

Given \gls{AoA} and \gls{RSS} extracted from the observed channel matrices, the detection problem aims to determine whether the measurements are legitimate or originate from a \gls{CSI} spoofing attack. We represent each measurement as a two-dimensional feature vector $\mathbf{x}(t) = [\hat{\theta}(t), \hat{P}(t)]^T$, where $\hat{\theta}(t)$ is the estimated \gls{AoA} and $\hat{P}(t)$ is the measured signal power. The detection task is formulated as distinguishing between two hypotheses: $\mathcal{H}_0$ representing legitimate transmissions drawn from the learned distribution $p_{\text{legit}}(\mathbf{x})$, and $\mathcal{H}_1$ representing spoofed transmissions from an unknown distribution $p_{\text{attack}}(\mathbf{x})$. The detection framework learns $p_{\text{legit}}(\mathbf{x})$ from attack-free training data and identifies deviations indicative of \gls{CSI} spoofing without prior knowledge of the specific attack trajectory or displacement. Detection performance is characterized by the \gls{FAR}, representing the fraction of legitimate data incorrectly flagged as attacks and the detection rate, representing the fraction of attack data correctly identified.

\section{Adaptive GMM-Based Spoofing Detection}\label{sec:gmm_detection}

The fundamental insight underlying our detection framework is that legitimate measurements exhibit consistent statistical patterns in the joint \gls{AoA}-\gls{RSS} feature space. These patterns emerge from physical constraints governing wireless propagation and user mobility. When an attacker injects \gls{CSI} from a spatially displaced location, the resulting measurements violate the learned distribution, creating detectable anomalies. The proposed adaptive \gls{GMM} framework learns these patterns from attack-free training data and identifies deviations without requiring prior knowledge.

\subsection{CSI Calibration and Feature Extraction}

Raw \gls{CSI} measurements are subject to hardware imperfections that must be calibrated before detection. Standard sequential calibration approaches \cite{delamouphysical} independently compensate for \gls{SFO}, \gls{STO}, \gls{CFO} and \gls{CPO} through linear regression and channel impulse response peak detection. However, these methods do not account for \gls{IQ} imbalance, characterized by gain mismatch $\varepsilon_g$ and phase error $\varepsilon_p$ between in-phase and quadrature components, the unknown time offset $\varsigma_t$, or antenna-specific phase errors $\xi_{\text{ant}}^{(m)}$ from RF chain variations.

In this work, we exploit a joint calibration framework adapted from \cite{9535488} that simultaneously estimates all impairment parameters through nonlinear least-squares optimization. Algorithm~\ref{alg:joint_calibration} summarizes the procedure: In Phase 1, we compute phase residuals between measured and ideal \gls{LoS} \gls{CSI} using ground-truth \gls{AoA}, Phase 2 decomposes residuals into frequency-domain impairments via nonlinear fitting, followed by Phase 3, which estimate per-antenna phase errors using circular mean to handle wraparound and finally, Phase 4 applies corrections to all measurements. This calibration significantly reduces the \gls{MAE} as shown in Table~\ref{tab:calibration_summary}, and is performed once during the initialization using approximately $N_{\text{cal}}$ reference measurements.

\begin{algorithm}[t]
    \caption{Phase and Antenna Calibration}
    \label{alg:joint_calibration}
    \scriptsize
    \begin{algorithmic}[1]
        \Require Raw CSI $\mathbf{H}_{\text{raw}}^{(i)} \in \mathbb{C}^{N_a \times N_\text{sc}}$ from $N_{\text{cal}}$ reference positions with known ground-truth AoA $\{\theta_{\text{gt}}^{(i)}\}_{i=1}^{N_{\text{cal}}}$
        \Ensure Calibrated CSI $\mathbf{H}_{\text{cal}}$
        \Statex \hrulefill
        \Statex \textbf{Phase 1: Compute Phase Residuals}
        \vspace{-0.55em}
        \Statex \hrulefill
        \For{$i = 1$ to $N_{\text{cal}}$}
            \For{$s = 1$ to $N_\text{sc}$}
                \State $\phi_{\text{ideal}}^{(i,s)} \gets -2\pi d \sin(\theta_{\text{gt}}^{(i)}) / \lambda$
                \State $\phi_{\text{raw}}^{(i,s)} \gets \text{mean}_m(\angle \mathbf{H}_{\text{raw}}^{(i)}[a,s])$
                \State $\Delta\phi^{(i,s)} \gets \phi_{\text{raw}}^{(i,s)} - \phi_{\text{ideal}}^{(i,s)}$
            \EndFor
        \EndFor
        \Statex \hrulefill
        \Statex \textbf{Phase 2: Frequency Domain Calibration}
        \vspace{-0.55em}
        \Statex \hrulefill
        \State $\{\varepsilon_g, \varepsilon_p, \varsigma_t, \zeta_{\text{SFO/STO}}, \eta_{\text{CPO}}\} \gets \arg\min_{\boldsymbol{\psi}} \sum_{i,s} |\Delta\phi^{(i,s)} - f(s; \boldsymbol{\psi})|^2$
        
        \Statex \hrulefill
        \Statex \textbf{Phase 3: Antenna Domain Calibration}
        \vspace{-0.55em}
        \Statex \hrulefill
        \For{$a = 1$ to $N_a$}
            \For{$i = 1$ to $N_{\text{cal}}$}
                \State $h_{\text{freq-corr}}^{(i,a)} \gets \frac{1}{N_\text{sc}}\sum_{s=1}^{N_\text{sc}} \mathbf{H}_{\text{raw}}^{(i)}[a,s] \cdot e^{-j\phi_{\text{freq}}^{(s)}}$
                \State $\Delta\phi_{\text{ant}}^{(i,a)} \gets \angle(h_{\text{freq-corr}}^{(i,a)}) - \phi_{\text{ideal}}^{(i)}$
            \EndFor
            \State $\xi_{\text{ant}}^{(a)} \gets \angle\left(\frac{1}{N_{\text{cal}}}\sum_{i=1}^{N_{\text{cal}}} e^{j\Delta\phi_{\text{ant}}^{(i,a)}}\right)$
        \EndFor
        
        \Statex \hrulefill
        \Statex \textbf{Phase 4: Apply Corrections}
        \vspace{-0.55em}
        \Statex \hrulefill
        \For{each measurement sample}
            \For{$a = 1$ to $N_a$, $s = 1$ to $N_\text{sc}$}
                \State $\mathbf{H}_{\text{cal}}[a,s] \gets \mathbf{H}_{\text{raw}}[a,s] \cdot e^{-j(\phi_{\text{freq}}^{(s)} + \xi_{\text{ant}}^{(a)})}$
            \EndFor
        \EndFor
        
        \State \Return $\mathbf{H}_{\text{cal}}$
    \end{algorithmic}
\end{algorithm}
\setlength{\textfloatsep}{0pt}

\begin{table}[htbp]
\centering
\caption{Calibration cac on measurement \gls{MAE}}
\label{tab:calibration_summary}
\begin{tabular}{lcccc}
\toprule
\multirow{2}{*}{Metric} 
    & \multicolumn{2}{c}{Uncalibrated} 
    & \multicolumn{2}{c}{Calibrated} \\
\cmidrule(lr){2-3} \cmidrule(lr){4-5}
    & Mean & Std & Mean & Std \\
\midrule
AoA ($^\circ$)   & 22.74 & 14.95 & 1.27 & 0.84 \\
RSS (dBm)        & 62.66 &  0.73 & 1.77 & 0.73 \\
\bottomrule
\end{tabular}
\end{table}
\setlength{\textfloatsep}{0pt}

\begin{figure}[htbp]
\centering
\begin{tikzpicture}
\begin{axis}[
    name=plot1,
    width=4.5cm,
    height=4cm,
    xlabel={AoA Error (degrees)},
    ylabel={CDF},
    xmin=0,
    ymin=0, ymax=1.05,
    grid=major,
    grid style={dashed, gray!30},
    legend pos=south east,
    legend style={font=\small, fill=white, fill opacity=0.9},
    tick label style={font=\small},
    label style={font=\normalsize},
]

\addplot[
    color=blue,
    line width=2pt,
    dashed,
] table[
    x=x,
    y=Uncalibrated,
    col sep=comma
] {csv/cdf_aoa.csv};
\addlegendentry{Uncal.}

\addplot[
    color=red,
    line width=2pt,
    solid,
] table[
    x=x,
    y=Calibrated,
    col sep=comma
] {csv/cdf_aoa.csv};
\addlegendentry{Cal.}
\end{axis}

\hspace{-4.5em}

\begin{axis}[
    at={(plot1.right of south east)},
    anchor=left of south west,
    xshift=1.5cm,
    width=4.5cm,
    height=4cm,
    xlabel={RSS Error (dBm)},
    ylabel={CDF},
    xmin=0,
    ymin=0, ymax=1.05,
    grid=major,
    grid style={dashed, gray!30},
    legend pos=south west,
    legend style={font=\small, fill=white, draw=black},
    tick label style={font=\small},
    label style={font=\normalsize},
]

\addplot[
    color=blue,
    line width=2pt,
    dashed,
] table[
    x=x,
    y=Uncalibrated,
    col sep=comma
] {csv/cdf_amplitude.csv};
\addlegendentry{Uncal.}

\addplot[
    color=red,
    line width=2pt,
    solid,
] table[
    x=x,
    y=Calibrated,
    col sep=comma
] {csv/cdf_amplitude.csv};
\addlegendentry{Cal.}
\end{axis}
\end{tikzpicture}
\caption{CDF of estimation \gls{MAE} before and after calibration of AoA and RSS.}
\label{fig:calibration_cdf}
\end{figure}
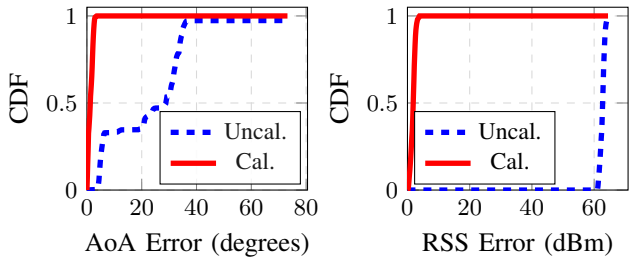
\setlength{\textfloatsep}{0pt}
From the calibrated channel matrix $\mathbf{H}_{\text{cal}}(t)$, we extract \gls{AoA} and \gls{RSS} using Root-MUSIC with spatial smoothing (subarray size of 16) and the average of received power across antennas and subcarriers, respectively \cite{8690541, delamouphysical}. While these metrics taken separately do not  determine a unique position, their joint distribution exhibits more precision and gives structured patterns during legitimate motion that are disrupted by \gls{CSI} spoofing attacks.

In addition, Fig.\ref{fig:calibration_cdf} illustrates the impact of calibration on measurement through \gls{CDF} of \gls{AoA} and amplitude \gls{MAE}. The left sub-figure shows that uncalibrated \gls{AoA} measurements exhibit substantial errors, with approximately 50\% of samples exceeding 25° deviation from ground truth. After calibration, the error distribution concentrates sharply near zero, with over 90\% of samples achieving errors below 3°. The right sub-figure demonstrates similar improvement for \gls{RSS}. This dramatic reduction in measurement errors ensures that the extracted features accurately capture the spatial characteristics necessary for distinguishing legitimate from spoofed \gls{CSI} patterns.

It is worth noting that we do not include \gls{ToF} in our feature set due to the limited delay resolution. With a bandwidth of $B=$ 20~MHz, the one-way theoretical delay resolution is \mbox{$\frac{1}{B}$ = 50~ns}, which corresponds to a spatial resolution of 15~m. Such a resolution can allow for only a coarse localization. In addition, the spatial resolution of the experimental dataset is 5~mm, which is much lower than the delay resolution. 

\subsection{Adaptive GMM Detection Model}
After calibration, we construct a two-dimensional feature vector at each time step $t$ as $\mathbf{x}(t) = [\theta_{\text{cal}}(t), P_{\text{cal}}(t)]^T$, representing the calibrated \gls{AoA} and \gls{RSS}. We model the probability distribution during legitimate operation using a \gls{GMM}, which provides a flexible probabilistic framework capable of capturing the natural variability and correlations in the joint feature space.

The \gls{GMM} density function for legitimate user $u$ is expressed as follows:
\begin{equation}
p_u(\mathbf{x}) = \sum_{k=1}^K \pi_k \mathcal{N}(\mathbf{x}; \boldsymbol{\mu}_k, \boldsymbol{\Sigma}_k)
\end{equation}
where $K$ represents the number of mixture components, $\pi_k$ denotes the mixing coefficients satisfying $\sum_k \pi_k = 1$, $\boldsymbol{\mu}_k \in \mathbb{R}^2$ represents the mean vector of component $k$ and $\boldsymbol{\Sigma}_k \in \mathbb{R}^{2\times2}$ denotes the full covariance matrix capturing correlations between \gls{AoA} and \gls{RSS}.

\subsubsection{Component Selection and Training}
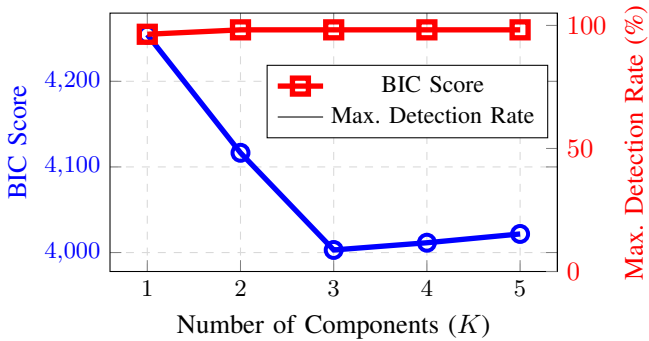
\begin{figure}[htbp]
\centering
\begin{tikzpicture}
\begin{axis}[
    name=plot1,
    width=7.5cm,
    height=5cm,
    xlabel={Number of Components ($K$)},
    ylabel={BIC Score},
    ylabel style={color=blue},
    yticklabel style={color=blue},
    xtick={1,2,3,4,5},
    grid=major,
    grid style={dashed, gray!30},
    tick label style={font=\small},
    label style={font=\normalsize},
]

\addplot[
    color=blue,
    line width=2pt,
    mark=o,
    mark size=3pt,
    mark options={fill=white, line width=1.5pt}
] table[
    x=k,
    y=bic,
    col sep=comma
] {csv/bic_detection_tradeoff.csv};

\end{axis}

\begin{axis}[
    at={(plot1.south west)},
    anchor=south west,
    width=7.5cm,
    height=5cm,
    xlabel={},
    ylabel={Max. Detection Rate (\%)},
    ylabel style={color=red},
    yticklabel style={color=red},
    axis y line*=right,
    axis x line=none,
    ymin=0, ymax=105,
    xtick={1,2,3,4,5},
    legend style={
        at={(0.35,0.8)},  
        anchor=north west,
        font=\small,
        fill=white,
        draw=black
    },
]

\addplot[
    color=red,
    line width=2pt,
    mark=square,
    mark size=3pt,
] table[
    x=k,
    y=max_detection,
    col sep=comma
] {csv/bic_detection_tradeoff.csv};

\addlegendimage{/pgfplots/refstyle=plot1}\addlegendentry{BIC Score}
\addlegendentry{Max. Detection Rate}

\end{axis}
\end{tikzpicture}
\caption{Model complexity trade-off: BIC score vs maximum detection performance across different numbers of GMM components. Lower BIC indicates better model fit, while higher detection rate shows superior attack detection capability.}
\label{fig:gmm_tradeoff}
\end{figure}

The number of components $K$ controls model complexity. We evaluate \mbox{$K \in \{1,2,3,4,5\}$} using both the Bayesian information criterion (BIC) and the maximum detection rate. We recall that the BIC is computed as $\text{BIC} = -2 \cdot \log \mathcal{L} + p_1 \cdot \log(p_2)$, where $\log \mathcal{L}$ is the log-likelihood of the model, $p_1 = (K-1) + K[d+\frac{p_3(p_3+1)}{2}]$ is the number of free parameters, $p_2$ is the number of samples and $p_3$ is the dimension or the number of features. Lower BIC values indicate better fitting. Fig.~\ref{fig:gmm_tradeoff} shows that $K=3$ components achieve the optimal balance, minimizing BIC while maximizing detection rate. This configuration provides sufficient flexibility to model the multi-modal structure of legitimate measurements without overfitting.

The \gls{GMM} parameters $\Theta = \{\pi_k, \boldsymbol{\mu}_k, \boldsymbol{\Sigma}_k\}_{k=1}^K$ are estimated using the expectation-maximization (EM) algorithm \cite{balakrishnan2017statistical, 543975}, which iteratively alternates between an expectation (E-step), which computes component responsibilities by evaluating the expected log-likelihood using current parameter estimates, and a maximization (M-step), which updates the GMM parameters to maximize this expected log-likelihood. This process continues until convergence on attack-free training data from multiple trajectories. Feature standardization is applied to ensure numerical stability: we compute the mean $\boldsymbol{\mu}_X$ and the standard deviation $\boldsymbol{\sigma}_X$ from the training set and apply $\mathbf{x}_{\text{scaled}} = (\mathbf{x} - \boldsymbol{\mu}_X) / \boldsymbol{\sigma}_X$ to all features before training.

\subsubsection{Adaptive Detection Framework}
Our framework incorporates two key enhancements beyond standard \gls{GMM}. First, exponential smoothing with factor $\delta$ is applied to the log-likelihood sequence to reduce sensitivity to instantaneous measurement fluctuations. Second, adaptive thresholding adjusts the detection threshold based on local window statistics, enabling robust performance.
The detector computes a windowed log-likelihood over $W$ samples as $\Lambda_u(t) = \sum_{i=t-W+1}^{t} \log p_u(\mathbf{x}_{\text{scaled}}(i); \Theta)$, which is smoothed as $\Lambda_{\text{smooth}}(t) =  \delta \Lambda(t) + (1-\delta) \Lambda_{\text{smooth}}(t-1)$.
The adaptive threshold function dynamically adjusts based on local statistics as follows:
\begin{equation}
\kappa_{\text{adapt}}(t) = \mu_{\text{local}}(t) - \beta \cdot \sigma_{\text{local}}(t),
\end{equation}
where $\mu_{\text{local}}(t)$ and $\sigma_{\text{local}}(t)$ are computed over an adaptive window size $W_{\text{adapt}}$ samples:
\begin{align}
\mu_{\text{local}}(t) &= \frac{1}{W_{\text{adapt}}} \sum_{i=t-W_{\text{adapt}}+1}^{t} \Lambda(i), \\
\sigma_{\text{local}}(t) &= \sqrt{\frac{1}{W_{\text{adapt}}} \sum_{i=t-W_{\text{adapt}}+1}^{t} (\Lambda(i) - \mu_{\text{local}}(t))^2},
\end{align}
with sensitivity parameter $\beta$. The adaptive threshold is constrained to be
\mbox{$\kappa_{\text{adapt}}(t) = \max(\kappa_{\text{adapt}}(t), 0.9 \cdot \kappa)$}, where $\kappa$ is the base threshold calibrated on validation data to achieve the target false alarm rate. An attack is declared when the exponentially smoothed likelihood $\Lambda_{\text{smooth}}(t)$ falls below $\kappa_{\text{adapt}}(t)$. Algorithm~\ref{alg:detection} summarizes the complete detection procedure, which processes measurements sequentially for real-time deployment. 

\begin{algorithm}[t]
\caption{Adaptive GMM Attack Detection}
\label{alg:detection}
\scriptsize
\begin{algorithmic}[1]
\Require Training data $\{\mathbf{x}^{\text{train}}(t)\}_{t=1}^{N_{\text{train}}}$, validation data $\{\mathbf{x}^{\text{val}}(t)\}_{t=1}^{N_{\text{val}}}$, test data $\{\mathbf{x}^{\text{test}}(t)\}_{t=1}^{N_{\text{test}}}$, number of components $K$, windows $W$ and $W_\text{adapt}$, smoothing factor $\delta$, convergence tolerance $\epsilon = 10^{-6}$, max iterations $T_{\text{max}} = 100$
\Ensure Binary vector $\{b(t)\}$

\Statex \hrulefill
\Statex \textbf{Phase 1: Feature Standardization}
\vspace{-0.55em}
\Statex \hrulefill
\State Compute $\boldsymbol{\mu}_X \gets \frac{1}{N_{\text{train}}} \sum_{t=1}^{N_{\text{train}}} \mathbf{x}^{\text{train}}(t)$
\State Compute $\boldsymbol{\sigma}_X \gets \sqrt{\frac{1}{N_{\text{train}}} \sum_{t=1}^{N_{\text{train}}} (\mathbf{x}^{\text{train}}(t) - \boldsymbol{\mu}_X)^2}$
\State Standardize all data: $\mathbf{x}_{\text{scaled}} \gets (\mathbf{x} - \boldsymbol{\mu}_X)/\boldsymbol{\sigma}_X$

\Statex \hrulefill
\Statex \textbf{Phase 2: GMM Training via EM Algorithm on the training data}
\vspace{-0.55em}
\Statex \hrulefill
\State Initialize parameters $\Theta^{(0)} = \{\pi_k^{(0)}, \boldsymbol{\mu}_k^{(0)}, \boldsymbol{\Sigma}_k^{(0)}\}_{k=1}^K$ randomly
\State $\mathcal{L}^{\text{new}} \gets -\infty$
\Repeat
    \State $\mathcal{L}^{\text{old}} \gets \mathcal{L}^{\text{new}}$  
    
    \State \Comment{E-step: Compute responsibilities}
    \For{$t = 1$ to $N_{\text{train}}$, $k = 1$ to $K$}
        \State $\gamma_k(t) \gets \frac{\pi_k \mathcal{N}(\mathbf{x}_{\text{scaled}}^{\text{train}}(t); \boldsymbol{\mu}_k, \boldsymbol{\Sigma}_k)}{\sum_{j=1}^{K} \pi_j \mathcal{N}(\mathbf{x}_{\text{scaled}}^{\text{train}}(t); \boldsymbol{\mu}_j, \boldsymbol{\Sigma}_j)}$
    \EndFor
    
    \State \Comment{M-step: Update parameters}
    \For{$k = 1$ to $K$}
        \State $N_k \gets \sum_{t=1}^{N_{\text{train}}} \gamma_k(t)$
        \State $\pi_k \gets N_k / N_{\text{train}}$
        \State $\boldsymbol{\mu}_k \gets \frac{1}{N_k} \sum_{t=1}^{N_{\text{train}}} \gamma_k(t) \mathbf{x}_{\text{scaled}}^{\text{train}}(t)$
        \State $\boldsymbol{\Sigma}_k \gets \frac{1}{N_k} \sum_{t=1}^{N_{\text{train}}} \gamma_k(t) (\mathbf{x}_{\text{scaled}}^{\text{train}}(t) - \boldsymbol{\mu}_k)(\mathbf{x}_{\text{scaled}}^{\text{train}}(t) - \boldsymbol{\mu}_k)^T$
    \EndFor
    
    \State Compute $\mathcal{L}^{\text{new}} \gets \sum_{t=1}^{N_{\text{train}}} \log \sum_{k=1}^{K} \pi_k \mathcal{N}(\mathbf{x}_{\text{scaled}}^{\text{train}}(t); \boldsymbol{\mu}_k, \boldsymbol{\Sigma}_k)$
\Until{$|\mathcal{L}^{\text{new}} - \mathcal{L}^{\text{old}}| < \epsilon$ or max iterations $T_{\text{max}}$ reached}

\Statex \hrulefill
\Statex \textbf{Phase 3: Threshold Calibration on the validation data}
\vspace{-0.55em}
\Statex \hrulefill
 
\For{$t = 1$ to $N_{\text{val}}$}
    \State Compute $\ell^{\text{val}}(t) \gets \log p_u(\mathbf{x}_{\text{scaled}}^{\text{val}}(t); \Theta)$
\EndFor
\For{$t = W$ to $N_{\text{val}}$}
    \State Compute windowed log-likelihood: $\Lambda^{\text{val}}(t) \gets \sum_{i=t-W+1}^{t} \ell^{\text{val}}(i)$
\EndFor
\State Set threshold $\kappa \gets$ 1st percentile of $\{\Lambda^{\text{val}}(t)\}_{t=W}^{N_{\text{val}}}$ to achieve 1\% FAR

\Statex \hrulefill
\Statex \textbf{Phase 4: Detection on test data}
\vspace{-0.55em}
\Statex \hrulefill
 
\For{$t = 1$ to $N_{\text{test}}$}
     \State Standardize: $\mathbf{x}_{\text{scaled}}^{\text{test}}$   \State Compute log-likelihood: $\ell(t) \gets \log p_u(\mathbf{x}_{\text{scaled}}^{\text{test}}(t); \Theta)$
\EndFor

\For{$t = W$ to $N_{\text{test}}$}
    \State Compute window log-likelihood: $\Lambda(t) \gets \sum_{i=t-W+1}^{t} \ell(i)$
    \State Apply exponential smoothing: $\Lambda_{\text{smooth}}(t) \gets \delta \Lambda(t) + (1-\delta) \Lambda_{\text{smooth}}(t-1)$
    \State Compute adaptive threshold: $\kappa_{\text{adapt}}(t)$
    \If{$\Lambda_{\text{smooth}}(t) < \kappa_{\text{adapt}}(t)$}
        \State $b(t) \gets 1$ \Comment{Attack detected}
    \Else
        \State $b(t) \gets 0$ \Comment{Legitimate}
    \EndIf
\EndFor

\State \Return $\{b(t)\}_{t=W}^{N_{\text{test}}}$
\end{algorithmic}
\end{algorithm}
\setlength{\textfloatsep}{0pt}

%
\section{Experimental Results}\label{sec:experiment_results}

\subsection{Experimental Setup}

Our experimental validation employs a massive MIMO testbed configured with a 64-element \gls{ULA}, deployed in a controlled laboratory environment with predominantly \gls{LoS} propagation conditions \cite{nr6k-8r78-21}. The dataset consists of channel measurements collected across parallel trajectories spaced 5~mm apart, with 251 measurement points per trajectory collected with a spacing of 5~mm. This ultra-dense spatial sampling enables precise evaluation of detection performance across a wide range of spatial displacements.

We partition the data into training (trajectories 1-50), validation (trajectories 51-60), and test (trajectories 65-70) subsets. The training set is used for the optimization of \gls{GMM} parameter, while the validation set calibrates the detection thresholds to achieve a target 1\% \gls{FAR}. The attack window spans samples 175-225 within the test trajectories, representing approximately 20\% of the trajectory length, ensuring sufficient legitimate data before and after the attack.\\
\indent We evaluate \gls{CSI} spoofing attacks by varying the spatial displacement from 5 mm to 450 mm, corresponding to trajectory offsets $o \in \{1, 2, \ldots, 90\}$ with each offset equal to 5~mm. Therefore, for each offset $o$, we transmit spoofed \gls{CSI} measurements from the trajectory at 5~mm $\times o$ during the attack window, simulating an adversary who has pre-recorded \gls{CSI} measurements from spatially displaced locations. The detection framework uses a \gls{GMM} with an optimal component number of $K=3$ as shown in Fig. \ref{fig:gmm_tradeoff}, $N_\text{cal} = 100$,  sliding windows $W = 10$, $W_{\text{adapt}} = 20$, and the weights are $\beta=1.5$, and $\delta=0.3$.

\subsection{Detection Performance and Analysis}

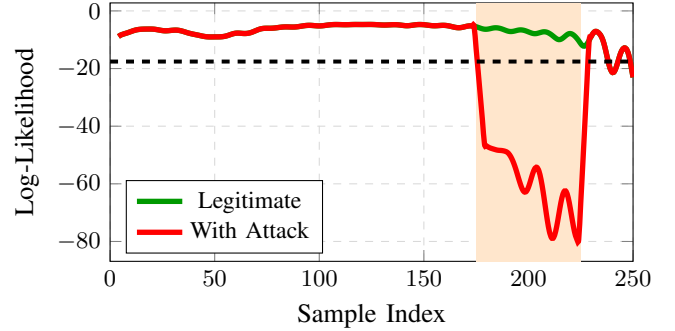
\begin{figure}[htbp]
    \centering
    \begin{tikzpicture}
    \begin{axis}[
        width=8.5cm,
        height=5cm,
        xlabel={Sample Index},
        ylabel={Log-Likelihood},
        xmin=0, xmax=250,
        grid=major,
        grid style={dashed, gray!30},
        legend pos=south west,
        legend style={font=\small, fill=white, draw=black},
        tick label style={font=\small},
        label style={font=\normalsize},
    ]
    
    \fill[orange!20] (axis cs:175,\pgfkeysvalueof{/pgfplots/ymin}) 
        rectangle (axis cs:225,\pgfkeysvalueof{/pgfplots/ymax});
    
    \addplot[
        color=green!60!black,
        line width=2pt,
        smooth
    ] table[
        x=Sample_Index,
        y=GMM_Lambda_Legit,
        col sep=comma,
        skip rows between index={0}{9}
    ] {csv/aoa/attack_detection_data_offset50.csv};
    \addlegendentry{Legitimate}
    
    \addplot[
        color=red,
        line width=2pt,
        smooth
    ] table[
        x=Sample_Index,
        y=GMM_Lambda_Attack,
        col sep=comma,
        skip rows between index={0}{9}
    ] {csv/aoa/attack_detection_data_offset50.csv};
    \addlegendentry{With Attack}
    
    \addplot[
        color=black,
        dashed,
        line width=1.5pt,
        forget plot
    ] table[
        x=Sample_Index,
        y=GMM_Threshold,
        col sep=comma,
        skip rows between index={0}{9}
    ] {csv/aoa/attack_detection_data_offset50.csv};
    \node[anchor=west, font=\small] at (axis cs:10,-220) {Threshold};
    
    \node[orange!80!black, font=\small\bfseries, anchor=north] 
        at (axis cs:200,\pgfkeysvalueof{/pgfplots/ymin}+2) {Attack Window};
    \end{axis}
    \end{tikzpicture}
    \caption{\gls{GMM} detection scores showing clear separation between legitimate and attack windows where the black dashed line is the threshold $\kappa_{\text{adapt}}$.}
    \label{fig:gmm_detection}
\end{figure}

Fig.~\ref{fig:gmm_detection} shows the adaptive \gls{GMM} detection score (windowed log-likelihood) for both legitimate and attacked trajectories. Before the attack, the log-likelihood remains consistently above the threshold. When the attack begins at sample index 175, the log-likelihood drops sharply below the threshold, triggering detection, demonstrating that the adaptive \gls{GMM} successfully learns the statistical structure of legitimate \gls{AoA}-\gls{RSS} patterns and reliably detects violations introduced by \gls{CSI} spoofing attack for a 250 mm spatial displacement.

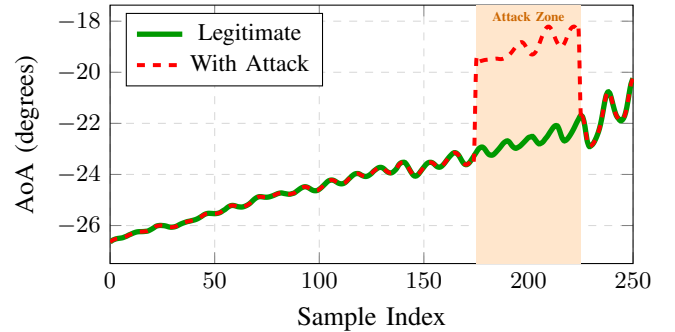
\begin{figure}[htbp]
    \centering
    \begin{tikzpicture}
    \begin{axis}[
        width=8.5cm,
        height=5cm,
        xlabel={Sample Index},
        ylabel={AoA (degrees)},
        xmin=0, xmax=250,
        grid=major,
        grid style={dashed, gray!30},
        legend pos=north west,
        legend style={font=\small, fill=white, draw=black},
        tick label style={font=\small},
        label style={font=\normalsize},
    ]
    
    \fill[orange!20] (axis cs:175,\pgfkeysvalueof{/pgfplots/ymin}) 
        rectangle (axis cs:225,\pgfkeysvalueof{/pgfplots/ymax});
    
    \addplot[
        color=green!60!black,
        line width=2pt,
        smooth
    ] table[
        x=Sample_Index,
        y=AoA_Legit,
        col sep=comma
    ] {csv/aoa/attack_detection_data_offset50.csv};
    \addlegendentry{Legitimate}
    
    \addplot[
        color=red,
        line width=1.5pt,
        dashed,
        smooth
    ] table[
        x=Sample_Index,
        y=AoA_Attack,
        col sep=comma
    ] {csv/aoa/attack_detection_data_offset50.csv};
    \addlegendentry{With Attack}
    
    \node[orange!80!black, font=\tiny\bfseries, anchor=south] 
        at (axis cs:200,\pgfkeysvalueof{/pgfplots/ymax}-1) {Attack Zone};
    \end{axis}
    \end{tikzpicture}
    \caption{\gls{AoA} measurements during \gls{CSI} spoofing attack with a displacement of 250 mm.}
    \label{fig:aoa_attack}
\end{figure}

\begin{figure}[htbp]
    \centering
    \begin{tikzpicture}
    \begin{axis}[
        width=8.5cm,
        height=5cm,
        xlabel={Sample Index},
        ylabel={Amplitude (dBm)},
        xmin=0, xmax=250,
        grid=major,
        grid style={dashed, gray!30},
        legend pos=south west,
        legend style={font=\small, fill=white, draw=black},
        tick label style={font=\small},
        label style={font=\normalsize},
    ]
    \fill[orange!20] (axis cs:175,\pgfkeysvalueof{/pgfplots/ymin}) 
        rectangle (axis cs:225,\pgfkeysvalueof{/pgfplots/ymax});
    
    \addplot[
        color=green!60!black,
        line width=2pt,
        smooth
    ] table[
        x=Sample_Index,
        y=Amp_Legit,
        col sep=comma
    ] {csv/aoa/attack_detection_data_offset50.csv};
    \addlegendentry{Legitimate}
    
    \addplot[
        color=red,
        line width=1.5pt,
        dashed,
        smooth
    ] table[
        x=Sample_Index,
        y=Amp_Attack,
        col sep=comma
    ] {csv/aoa/attack_detection_data_offset50.csv};
    \addlegendentry{With Attack}
    
    \node[orange!80!black, font=\tiny\bfseries, anchor=south] 
        at (axis cs:200,\pgfkeysvalueof{/pgfplots/ymax}-1) {Attack Zone};
    \end{axis}
    \end{tikzpicture}
    \caption{Signal amplitude during \gls{CSI} spoofing attack with a displacement of 250 mm.}
    \label{fig:amplitude_attack}
\end{figure}
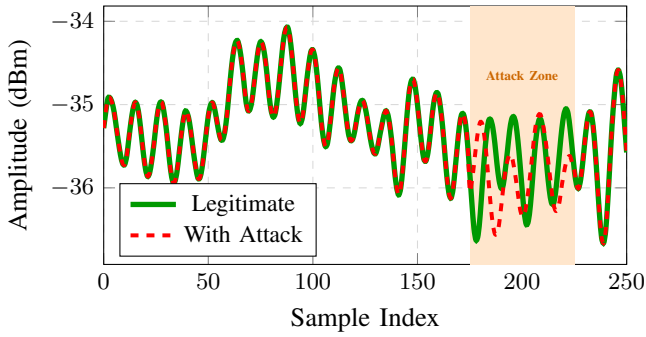

Fig.~\ref{fig:aoa_attack} illustrates the effect of \gls{CSI} spoofing on \gls{AoA}. During the attack, the spoofed \glspl{CSI} deviate significantly from the legitimate trajectory. Similarly, Fig.~\ref{fig:amplitude_attack} shows the impact on \gls{RSS}. The joint \gls{AoA}-\gls{RSS} distribution exhibits clear deviations from legitimate pattern, demonstrating that it creates a highly detectable anomaly.

\begin{figure}[htbp]
\centering
\begin{tikzpicture}
\begin{axis}[
    width=8.5cm,
    height=4.5cm,
    xlabel={Spatial Displacement (mm)},
    ylabel={Detection Rate (\%)},
    xmin=0, xmax=450,
    ymin=0, ymax=105,
    grid=major,
    grid style={dashed, gray!30},
    legend pos=south east,
   legend style={font=\small, fill=white, draw=black},
    tick label style={font=\small},
    ymajorgrids=true,
    xmajorgrids=true,
    xtick={100,200,300,400,500},
    extra x ticks={0},
    extra x tick labels={},
]
\addplot[
    dashed,
    gray,
    line width=1.5pt,
    forget plot
] coordinates {(0,90) (500,90)};
\node[gray, anchor=west] at (axis cs:50,85) {\small 90\%};

\addplot[
    color=green!60!black,
    mark=o,
    mark size=3pt,
    line width=2.5pt,
] table[
    x=Displacement_mm,
    y=GMM_Pure_DR,
    col sep=comma
] {csv/aoa/resolution_analysis.csv};
\addlegendentry{GMM}

\addplot[
    color=green!40!black,
    mark=+,
    mark size=3pt,
    line width=2.5pt,
] table[
    x=Displacement_mm,
    y=GMM_Enhanced_DR,
    col sep=comma
] {csv/aoa/resolution_analysis.csv};
\addlegendentry{Adaptive GMM}

\addplot[
    color=red,
    mark=square*,
    mark size=2.5pt,
    line width=2pt,
] table[
    x=Displacement_mm,
    y=LOF_DR,
    col sep=comma
] {csv/aoa/resolution_analysis.csv};
\addlegendentry{LOF}

\addplot[
    color=blue,
    mark=triangle*,
    mark size=2.5pt,
    line width=2pt,
] table[
    x=Displacement_mm,
    y=SVM_DR,
    col sep=comma
] {csv/aoa/resolution_analysis.csv};
\addlegendentry{SVM}

\addplot[
    color=orange,
    mark=diamond*,
    mark size=3pt,
    line width=2pt,
] table[
    x=Displacement_mm,
    y=kNN_DR,
    col sep=comma
] {csv/aoa/resolution_analysis.csv};
\addlegendentry{k-NN}

\end{axis}
\end{tikzpicture}
\caption{Detection rate versus spatial displacement for all evaluated algorithms.}
\label{fig:detection_vs_displacement}
\end{figure}
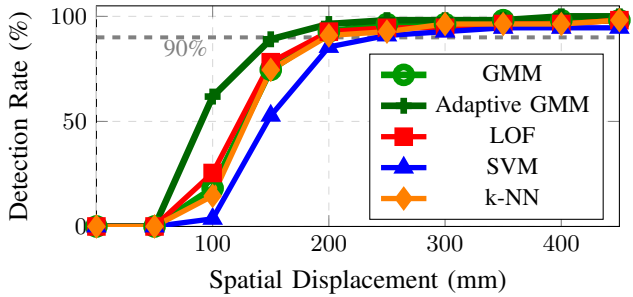

Fig.~\ref{fig:detection_vs_displacement} depicts detection performance as a function of spatial displacement. For displacements above 300 mm, all the algorithms can achieve a detection rate exceeding 95\%, demonstrating robust performance for large spatial offsets. The critical displacement range lies between 100-250 mm, where the anomalies become difficult to detect as the spoofed \gls{CSI} closely resembles legitimate patterns. Notably, adaptive \gls{GMM} demonstrates superior performance in the low-displacement regime, achieving a datection rate of 70\% at 100 mm compared to 5\%, 20\%, 22\% and 25\% 
for \gls{SVM}, \gls{k-NN}, conventional \gls{GMM} and \gls{LOF}, respectively.

\section{Conclusion}\label{sec:conclusion}

We proposed an adaptive \gls{GMM}-based framework for detecting \gls{CSI} spoofing attacks in massive MIMO wireless localization systems through joint analysis of calibrated \gls{AoA} and \gls{RSS} measurements. Comprehensive calibration reduces the \gls{MAE} of \gls{AoA} and \gls{RSS} from 22.74° to 1.27° and from 62.66 dBm to 1.77 dBm, respectively, enabling highly accurate estimation of the those parameters. The validation on a the dataset demonstrates superior detection rate with 1\% \gls{FAR}, using our proposed adaptive GMM compared to baseline methods such as conventional \gls{GMM}, \gls{LOF}, \gls{SVM} and k-NN. Future work will investigate extensions to non-line-of-sight multipath environments and integration with complementary security mechanisms.

\section*{Acknowledgement}
This work was partially funded by Horizon Europe SNS JU ROBUST-6G project under Grant Agreement No. 101139068, EU HORIZON-MSCA-SE TRACE-V2X project under Grant Agreement No. 101131204, and French National Research Agency under France 2030 program PEPR Networks of the Future Projects (NF-YACARI: ANR-22-PEFT-0005, NF-HiSec: ANR-22-PEFT-0008, NF-FOUNDS: ANR-22-PEFT-0009), ANR JCJC SecLoc project (ANR-24-CE45-3795), as well as the TalCyb Chair on Cybersecurity.

\balance
\bibliographystyle{IEEEtran}
\bibliography{bibliography}
\end{document}